\documentclass[twocolumn,showpacs,preprintnumbers,amsmath,amssymb]{revtex4}

\usepackage{graphicx}
\usepackage{dcolumn}
\usepackage{bm}

\begin{document}

\title{Field-Induced Insulator to Semimetal Transition and Field
Electron Emission of Nanorods of Semiconductors of Wide Energy
Band Gaps}

\author{Zhibing Li, Weiliang Wang, Shaozhi Deng, Ningsheng
Xu\footnote{corresponding author, stsxns@zsu.edu.cn} }
\affiliation{The State Key Laboratory of Optoelectronic Materials
and Technologies Department of Physics, Zhongshan University,
Guangzhou, 510275, China }
\author{Guiyang Huang}
\affiliation{Department of Physics, Tsinghua University, Beijing,
China }

\vspace{1pt}

\begin{abstract} Significant field emission is found theoretically
possible from nanorods of semiconductors of wide energy band gaps.
If the nanorod has a thin surface layer containing a large number of
localized states, a part of nanorod can exhibit an insulator-to-semimetal
transition under high enough fields of direction parallel to its axis, so
that field emission occurs at the apex of the metal like tip. The field
emission property of silicon carbide nanorods is studied as an example
and found to be in qualitative agreement with the experimental findings.

Key words: field emission; nanorod; silicon carbide

\end{abstract}
\vspace{1pt}

\pacs{85.45.Db,\ 72.20.Ee,\ 73.23.-b,\ 79.70.+q}

\vspace{1pt}
\maketitle

\vspace{1pt}

Quasi one-dimensional materials have attracted great interest
because of their unusual electronic and mechanical properties.
Since the discovery of carbon nanotubes [1], nanorods of many
materials have been synthesized, including some wide energy band
gap semiconductors, such as GaN [2,3] and SiC [4,5,6,7]. The wide
band gap semiconductors offer the potential to develop a new
generation of micro- and nano-electronic devices that can operate
at high power levels, high temperatures, and in harsh environment
[8,9,10,11]. They are attractive also because they usually have
large thermal conductivity and high saturation drift velocity.
Recently, it is reported that the SiC nanorods exhibit high
electron field emission with high stability [4,5,12], in
particular, under the lowest turn-on and threshold fields so far
reported for nanowires and nanorods [4,5]. However, the physical
mechanism responsible for field emission from quasi
one-dimensional wide band gap semiconductor materials is not
clear, for example, the physical origins of field enhancement at
and the continuing supply of electrons to an emission site. In
this letter£¬we propose a mechanism for explaining such a
phenomenon.

The bulk wide band gap semiconductors without doping are actually
insulators at room temperature. When they are made into nanorods,
which have high aspect ratios, the densities of localized states
induced by the atoms in the surface layer are high. This fact
enables that electrons transport via hopping from a metal
substrate to the apex of the free tip end of the nanorod when an
external field is applied parallel to its axis. In the following,
we show several physical processes indicated below will occur:
When charges induced by the applied field raise the Fermi level
close to the bottom of conduction band (BOCB), an insulator to
semimetal transition may occur. Then the work function is replaced
by the affinity as the vacuum barrier height. The latter is much
smaller than the former. At the same time, the degenerate
electrons in the metal-like region shield the field and lead to
field enhancement at the tip apex.

The set-up for field emission under consideration is schematically
illustrated in Fig.1a, the cathode and the anode made of metal is
parallel to each other and separated with a large enough distance.
A columniform nanorod of radius $r_{0}$ and with a spherical apex
is mounted vertically on the cathode and forms a semiconductor and
metal contact (we call it back contact) with the cathode.

\begin{figure}
\includegraphics{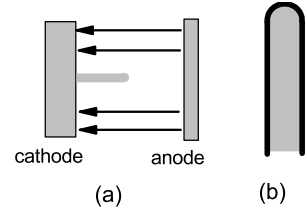} \caption{\label{fig:1}  (a) Set-up
for field emission. A nanorod is mounted on the metal cathode
vertically. (b) The nanorod has a radius of $r_{0}$. The surface
states are located in the surface layer (the dark line) of
thickness $\lambda$.}
\end{figure}

When an electric field is initially applied to the nanorod, the
BOCB along the nanorod decreases due to field penetration, so
negative charges will move to and accumulate in the tip-region of
the nanorod. The continuous supply of negative charges at this
stage is by electron hopping through the back contact from
substrate. The Coulomb potential of these negative charges will
stop BOCB to decrease farther. The balance between charges and
field follows the Poisson equation.

The Fermi level of the nanorod will be continuous at the back
contact and keeps constant with that of the metal substrate, but
it will vary along the nanorod if there is an electric current.
For wide band gap bulk materials of no doping, the electric
current is negligible. The case for nanorods could be much
different since there are many states due to the existence of
defects in the surface layer of the nanorods. A recent careful
simulation to the Si nanowire has revealed semimetal and metal
states originating in the dimer dangling bonds and surface
reconstruction [13]. The Shockley model suggests that the number
of surface states has the same order of surface atoms [14]. We may
assume that this model applies to the case of a surface layer full
of defects. It is also known that the surface states and defect
states are often inside the band gap and localized. In our model,
the area density of localized states due to the surface layer
$g_{s}(E,E_{c})=g_{s}(E-E_{c})$ is nonzero only for
$-E_{s}<E-E_{c}<0$, with $E_{c}$ the BOCB. The integral $\int
g_{s}(E-E_{s})dE=\sigma$ is a constant proportional to the number
of surface atoms per area. A localization parameter $\lambda$ is
defined to be the average spatial extension of localized states.
Then the charges of electrons in localized states distribute in a
surface layer of thickness $\lambda$ (Fig. 1b). The electric
current is given by
\begin{eqnarray}
J&=&-\pi r_{0}^{2} n \mu_{n} \frac{dE_{f}}{dz} - 2\pi e r_{0} R
g_{sf} kT \nu_{ph}\cdot \nonumber \\
&~& \exp\left(-\frac{2R}{\lambda}-\frac{\Delta W}{kT}\right)
\sinh\left(-\frac{R}{kT} \frac{dE_{f}}{dz}\right)
\end{eqnarray}
where the first term is the current of free electrons in the
conduction band, while the second term is the contribution of the
hopping current [15] that exists only if the Fermi level is inside
the band gap, $n$ and $\nu_{n}$ respectively the density and the
mobility of conducting electrons, $E_{f}$ the Fermi level,
$g_{sf}=g_{s}(E_{f},E_{c})$ the density of localized states with
Fermi energy, $R=\left(\frac{\lambda}{g_{sf}2\pi kT}\right)^{1/3}$
the mean range of hopping, $\nu_{ph}$ the frequency of phonons,
$\Delta W=\frac{1}{g_{sf} 2\pi R^{2}}$ the mean energy difference
between localized states of electrons before and after hopping.

The electrons occupy the localized states and conduction band
states according to the Fermi-Dirac distribution. The charge
density inside the nanorod is \begin{equation}
\rho(\textbf{r})=\rho_{s}(\textbf{r)}+\rho_{c}(\textbf{r})
\end{equation}
The density of charges in localized states, $\rho_{s}(\textbf{r})$
, is zero outside the surface layer, while inside it is give
\begin{equation}
\rho_{s}=\frac{-e}{\lambda}\int^{E_{c}}_{E_{f}^{*}}
g_{s}(E-E_{s})\frac{2}{\exp(\beta(E-E_{f}))+1}dE
\end{equation}
where $E_{f}^{*}$  is the intrinsic Fermi level of the neutral
nanorod. The density of charges in the conduction band is
\begin{equation}
\rho_{c}(\textbf{r})=-e\int^{\infty}_{E_{c}}g_{c}(E-E_{c})\frac{2}{\exp(\beta
(E-E_{f}))+1}dE \end{equation} where $g_{c}(E-E_{c})$ is the
density of states of conduction electrons. For simplicity, we will
assume that the localized states have a uniform energy
distribution in the band gap, i.e., the density of localized
states
\begin{equation}
g_{s}(E-E_{c})=\frac{\sigma}{E_{g}}\theta(E-E_{c}+E_{g})\theta(E_{c}-E)
\end{equation}

For a given $\rho(\textbf{r})$ , the electrostatic potential is
the solution of Poisson equation satisfying certain boundary
conditions. However, the charge density also depends on the
potential through the position of the BOCB and Fermi level. The
latter is related to the electric current, which is a constant
along the nanorod and equals to the emission current. These
entangled quantities are calculated by an iteration process.
Firstly, with a given function of Fermi level (initially it is
aligned with the Fermi level of substrate), the charge density is
given by Eqs.(2)-(4). Secondly the electrostatic potential is
calculated by solving the Poisson equation. The third step is to
calculate the emission current in the WKB approximation. The forth
step is to solve Eq. (1) to obtain a new function of Fermi level.
This process is repeated until a satisfactory consistency is
reached.

Now we take the wurtzite polytype of SiC as an example to
demonstrate our calculation. The band gap is 3.0 $eV$, the work
function 4.4 $eV$, the dielectric constant 10.32, and the mobility
of electrons 300. $cm^{2}/Vs$ [16]. The length of nanorod is
assumed to be 2.0 $\mu m$. The radius is 10.0 $nm$ if is not
stated otherwise. The density of localized states is assumed to be
$2.\cdot 10^{12}/cm^{2}$, while $\lambda$ is assumed to be 2.0
$nm$ which resembles to that of Si surface [17]. The work function
of substrate (assuming to be tungsten, W) is 4.58 $eV$. The value
of $\nu_{ph}$ is $2.4\cdot 10^{13} s^{-1}$. As a step of
iteration, the electrostatic potential is obtained by employing
finite element method for a given charge distribution. Fig. 2 is
an instance of 8.0 $V/\mu m$. The vertical axis is also the axis
of the nanorod. In the upper region of the nanorod, the field is
shielded as that part of the nanorod is metal-like. Due to field
enhancement, the potential barrier in front of the tip apex is the
thinnest (Fig. 2b). Thus, only field emission from the tip apex is
considered below.

\begin{figure}
\includegraphics{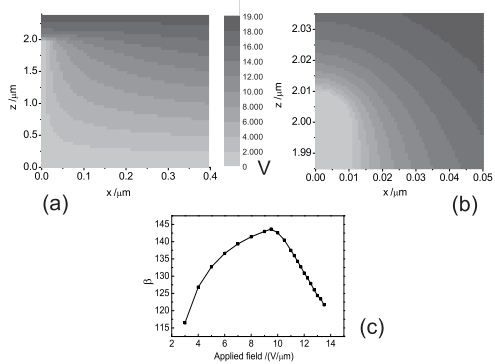}
\caption{\label{fig:2} (a) Illustrating electrical potential when
applied field is $8.0 V/\mu m$. The nanorod (radius $10.0 nm$,
length $2.01 \mu m$) is on Y axis. The electrostatic field is not
shielded at the bottom of the nanorod, while at the tip apex the
electrostatic field is almost completely shielded. (b) Details in
the tip region. (c) The field enhancement factor $\beta$ changes
with applied field nonlinearly. }
\end{figure}

The calculated results of the BOCB (solid line) and Fermi level
(dotted line) along the nanorod of diameter of 10.0 $nm$ are shown
in Fig. 3a. Under the applied field of 3.0 $V/\mu m$, the entire
nanorod is an insulator. Under 5.0 $V/\mu m$, there is a region
including the tip where the BOCB is quite close to the Fermi
level. Under 10.5 $V/\mu m$, the region is wider, where the BOCB
almost coincides with the Fermi level. This region is the
metal-like region. When the applied field is smaller than 10.5
$V/\mu m$, the emission current is tiny, so is the current in the
nanorod. Thus the Fermi level is almost unchanged. When the
applied field is greater and the electric current cannot be
ignored, the Fermi level begins to decrease. The BOCB decreases
linearly until the number of electrons in the conduction band is
large enough to stop the decrease of BOCB as the BOCB approaches
to the Fermi level. In the metal-like region the current is
provided by the electrons in the conduction band (the first term
of Eq. (1) ). Since the density of electrons in the conduction
band, n, is large in this region, the slop of Fermi level must
approach zero in order to keep a constant current. The Fermi level
obtained from Eq. (1) is basically horizontal in the metal-like
region.

The above typical feature is due to the effect of nano-dimension.
Taking 8.0 $V/\mu m$ as an instance, the dependence of band
diagram on the radius of nanorod is shown in Fig. 3b, revealing
that when the radius is larger than 40.0 $nm$, the metal-like
region disappears. This can explain why needle-shaped SiC nanorods
[12] and nano protrusions on the tip apex of a ZnO nanorod [18]
can improve field emission. The above effect may be understood as
follows. When a region becomes metal-like, it then has a high
density of electrons in conduction band. Thus, we have to explain
how under high fields a section of a nanorod can have a high
density of electrons in conduction band. Before the fields are
high enough to cause electron emission into vacuum, under the
action of applied fields, electrons will move to the tip region
through hopping in the surface layer, where the electrons will be
trapped in the localized states, which occupy the band gap of the
surface layer. As the applied field increases, more and more
electrons come to the tip region but cannot yet emit into vacuum.
Thus they accumulate in the region so that the localized states
will be gradually filled up. Accordingly, the Fermi level
gradually approaches the BOCB. Under room temperature, those
electrons in the states above Fermi level can hop to conduction
band when the energy difference between the Fermi level and the
BOCB is small. All of the electrons in conduction band are not
localized; i.e. they can move to the core region of the nanorod.
This situation remains if field electron emission occurs; it is
difficult for electrons in the localized states to be emitted
since surface barrier is too high. But why can this only happen in
small nanorods? The density mentioned above can be high only when
the density of conduction electrons in the surface layer is very
high, and only when the ratio of the volume of the surface layer
over that of core region of the nanorod is large enough. The
thickness of the surface layer is often fixed, so the second ratio
will be low if the core region is too large. In this case, the
insulator-semimetal transition will not occur. This explains why a
nanorod of large diameter has no metal-like region.

\begin{figure}
\includegraphics{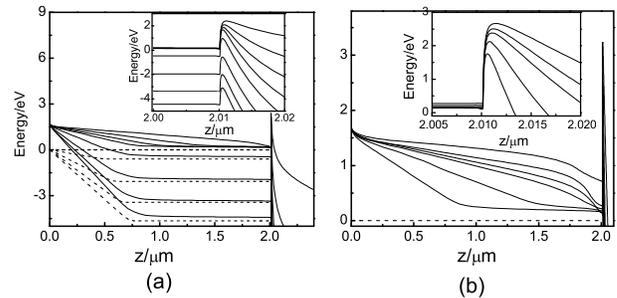}
\caption{\label{fig:3}  (a) The solid lines in range 0 to 2.01
$\mu m$ are BOCBs. The lines from up to down are for applied
fields 3.0, 5.0, 7.0, 9.0, 10.5, 11.75, 12.75, and 13.5 $V/\mu m$
respectively. Those in $z > 2.01 \mu m$ are the vacuum barriers
under various fields. The dashed lines are the Fermi levels. The
Fermi levels for four smaller fields are very close to zero. For
fields greater than $9.0 V/\mu m$ the bending of Fermi level is
significant. (b) The potential diagram in a field of 8.0 $V/\mu m$
for SiC nanorods of various radii. From up to down, the radii are
200, 100, 60.0, 40.0, 20.0, and 10.0 nm, respectively. The
horizontal dashed line is the Fermi level.}
\end{figure}

When the tip region becomes metal-like, field enhancement at the
apex surface appears, and become more and more obvious while
getting more and more metallic (Fig. 2(c)). However, the
enhancement factor decreases when the applied field is higher than
certain value (it is incidentally the turn-on field). That is
because the current is significantly high, leading to a large
voltage drop in the section of nanorod of not being metal-like,
i.e., the effective voltage across the vacuum gap becomes smaller.

This is more clear from the following analysis. In the insulating
region, the hopping current is dominated, while in the metal-like
region, the conduction current is dominated. Taking 13.5 $V/\mu m$
as an example, the hopping current (the second term of Eq.(1)) and
conduction current (the first term of Eq.(1)) are presented in
Fig. 4(a).

\begin{figure}
\includegraphics{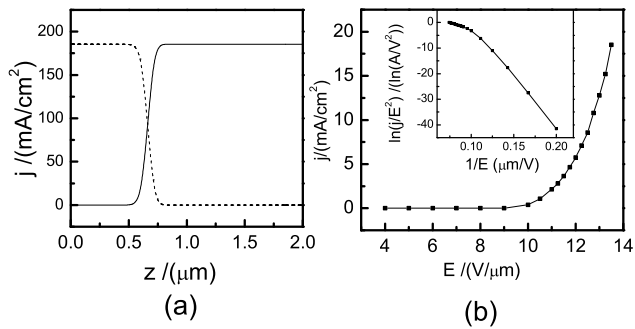}
\caption{\label{fig:4} The hopping current density (dashed) and
the conduction current density (solid) along the SiC nanorod of
length 2.01 $\mu m$ and radius 10.0 nm. The appliced field is 13.5
$V/\mu m$. (b)The emission current density versus the applied
field. The inset is the corresponding FN plot. }
\end{figure}

The field emission characteristics are obtained by assuming
Fowler-Nordheim tunneling [19]. Fig. 4(b) shows the field emission
current density versus applied field (the $J-E$ curve, with
current density $J$ defined by the emission current divided by the
cross section area of the nanorod). The inset is the FN plot ($\ln
(J/E^{2})$ versus $1/E$ ). For the fields larger than the turn-on
field, the slop of the FN plot decreases, as observed in
experiment [4,5,12]. This nonlinear behavior of the FN plot is
different from that predicted for a metal surface by the classical
theory, i.e., it should be a straight line.

The field emission energy distribution (FEED) is given in Fig.
5(a) for 13.25 $V/\mu m$ and (b) for 10.0 $V/\mu m$. The peak of
the FEED shifts toward lower energy while the applied field
increases, indicating existence of potential drop in the nanorod.
Similar phenomenon was also observed experimentally for
non-metallic emission regimes [20,21], but have not yet been
reported for nanorods and nanowires. Our calculation shows that
the FEED spectrum is sharply cut by the BOCB, only tiny
contribution from the localized states is observed in Fig.5(a).

\begin{figure}
\includegraphics{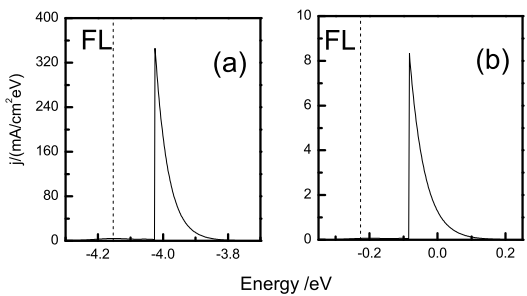}
\caption{\label{fig:5} The FEEDs for applied field 13.25 $V/\mu m$
(a) and 10.0 $V/\mu m$ (b) respectively. The peak shifts toward
lower energy while the applied field increases. The bottom of
conduction band makes the sharp cuts of the low energy tails. The
dashed lines are the Fermi levels at the end of the nanorod for
the corresponding applied fields. }
\end{figure}

To conclude, a mechanism for field emission from nanorods of wide
band gap semiconductors is proposed. As an example, the band
diagrams and vacuum barriers of SiC nanorods are calculated. A
field-induced insulator to semimetal transition is shown. This
transition is responsible to the efficient field emission from the
nanorods. It is found that the field enhancement is not a constant
with applied fields but varies with the field and has a maximum.
The FN plot of the emission current is apparently different from
straight line. The field emission energy distribution shows the
lowering of Fermi level as applied field increases. We should
stress that the appearance of the metal-like region where
electrons become degenerate is only possible in the nanoscale.

Authors gratefully acknowledge financial support of the projects
from the National Natural Science Foundation of China
(Distinguished Creative Group Project; Grant No. 90103028,
90306016), from the Education Ministry of China, from the Higher
Education Bureau, and from the Science and Technology Commssion of
Guangdong Province.

\end{document}